\documentclass[%
 reprint,
showpacs,preprintnumbers,
 amsmath,amssymb,
aps,
pra,
]{revtex4-1}
\usepackage{graphicx}% Include figure files
\usepackage{dcolumn}% Align table columns on decimal point
\usepackage{bm}% bold math
\begin{document}
\title{Unified Theoretical Approach to Electronic Transport from Diffusive to Ballistic
Regimes}% Force line breaks with \\
\author{H. Geng$^1$}
\author{W. Y. Deng$^1$}
\author{Y. J. Ren$^1$}
\author{L. Sheng$^{1,2}$}
\email{shengli@nju.edu.cn}
\author{D. Y. Xing$^{1,2}$}%
\affiliation{
 $^1$ National Laboratory of Solid State Microstructures and Department of Physics, Nanjing University, Nanjing 210093, China\\
 $^2$ Collaborative Innovation Center of Advanced Microstructures,
Nanjing University, Nanjing 210093, China
}
\begin{abstract}
We show that by integrating out the electric field
and incorporating proper boundary conditions, a semiclassical Boltzmann equation can
describe electron transport properties, continuously from the diffusive to ballistic regimes.
General analytical formulas of the conductance in $D=1,2,3$ dimensions are obtained,
which recover the Boltzmann-Drude formula and Landauer-B\"{u}ttiker formula in the 
diffusive and ballistic limits, respectively.
This intuitive and efficient approach can be applied to investigate the interplay of
system size and impurity scattering in various charge and spin transport phenomena.
\end{abstract}


\maketitle

\section{Introduction}\label{s1}
%Ò»°ãÇé¿ö
The Boltzmann equation was first devised by Ludwig Boltzmann in 1872
to describe the state of a dilute gas~\cite{BoltzOrg}.
In the modern literature the term Boltzmann equation often
refers to any kinetic equation that describes the change of a macroscopic quantity in a
nonequilibrium thermodynamic system, such as energy, charge or particle number.
The Boltzmann equation has proven fruitful not only for
the study of the classical gases, but also, properly generalized,
for electron transport in nuclear reactors, photon transport
in superfluids, and radiative transport in
planetary and stellar atmospheres~\cite{BltzFruits}. In condensed matter physics,
among many successes,
an important achievement based upon the Boltzmann equation is
the Drude kinetic theory of electrical conduction, which was proposed in
1900 by Paul Drude to explain the transport properties of electrons in
macroscopic conductors~\cite{Drude}.
The Boltzmann-Drude  formula for the zero-frequency conductance
of a conductor with length $L_{x}$ and cross section $A$ is
\begin{equation}\label{BD}
G=\frac{A}{L_x}\frac{n_e e^2\tau_0}{m_e}\ ,
\end{equation}
which correctly relates the conductance to the electron density $n_e$ 
and the relaxation time $\tau_0$ due to electron scattering by
impurities. It works very well for macroscopic conductors, where 
the sample size is much greater than the electron mean free path
(diffusive regime). 
Moreover, this elegant formula can be reproduced by using
modern linear-response theory and Green's function technique
through a deliberate summation of an infinite series of ladder diagrams~\cite{Mahan}.

Mesoscopic systems have been subject to
tremendous investigations in recent years. Theoretical works
on the electronic transport properties of mesoscopic systems
are often based upon the transmission approach. In this approach,
a conductor is viewed as a target, at which the incident carriers
are reflected or transmitted to other probes.
The (two-terminal) conductance of a conductor is given by
the famous Landauer-B\"{u}ttiker formula~\cite{LB1,LB2}
\begin{equation}\label{Landau}
G=\frac{e^2}{h}\sum_{n=1}^{N_{ch}}T_n\ ,
\end{equation}
where $T_{n}$ is the transmission cofficient of the $n$-th conducting
channel, and $N_{\mathrm{ch}}$ is the total number of conducting channels.
The Landauer-B\"{u}ttiker formula was originally proposed
based upon phenomenological discussions~\cite{LB1,LB2}, and later shown to 
be equivalent to the Kubo linear-response theory~\cite{LB_Ln1,LB_Ln2,LB_Ln3}.

The size of a mesoscopic conductor can be much smaller than the
electron mean free path, and it may even
be free of impurities. The Boltzmann-Drude formula
fails to behave properly in this ballistic regime.
According to the Landauer-B\"{u}ttiker
formula, $T_{n}\rightarrow 1$ in this regime, and the conductance saturates to a finite
value $G=\frac{e^2}{h}N_{\mathrm{ch}}$. In contrast, the Boltzmann-Drude
formula diverges with vanishing impurity scattering ($\tau_{0}\rightarrow\infty$). 
In the opposite diffusive regime, 
in principle both formulas, Eqs.\ (\ref{BD}) and (\ref{Landau}), should be applicable, but  
calculations using the Landauer-B\"{u}ttiker formula 
are generally very difficult and impracticable. Research works~\cite{Fenton,Pasta}
were devoted to developing unified theories covering both regimes. However, lengthy
sophisticated Green's function calculations were performed,
and concrete results were obtained only in some special limiting cases,
which make the theories hardly useful in practice.
Practicable and intuitive electronic transport theory, which can seamlessly
bridge the diffusive and ballistic regimes, is still awaited.

In this paper, we show that by integrating out the position-dependent electric field
and incorporating proper boundary conditions, a Boltzmann equation can
describe electron transport properties of a finite-size conductor, continuously from the
diffusive to ballistic regimes. We present both exact numerical and approximate analytical
solutions to the Boltzmann equation. General analytical formulas of the
conductance in $D=1,2,3$ dimensions are obtained, which are consistent with
the Boltzmann-Drude formula in the diffusive regime, and recover the Landauer-B\"{u}ttiker
formula in the ballistic regime. The theory has the advantage of being
simple and intuitive, and can be applied to study
the interplay of system size and impurity scattering in various charge and spin
transport phenomena.

In the next section, we introduce the Boltzmann equation and associated boundary conditions
in our model.
Through a linear transformation, the electric field with unknown position dependence is integrated
out from the Boltzmann equation. In Sec.\ \ref{s21}, the exact solution of this model is
obtained numerically. In Sec.\ \ref{s3}, an analytical approximate theory is developed,
and analytical formulas of the conductance in different dimensions are obtained.
The final section contains a summary and some discussions.

\section{Model Description}\label{s2}

Let us consider a finite-size conductor of dimension $D$
($D=1$, $2$, or $3$). The classical nonequibrium
distribution function $f({\bf k},{\bf r})$ of the electrons in the sample is
a function of the phase space point $({\bf k},{\bf r})$ with ${\bf k}$ and ${\bf r}$
as the momentum and coordinate of an electron. It is assumed that the quantum phase coherence
length $l_{\varphi}$ is smaller than the electron mean free path $l_{f}$,
such that the interference of electron
scattering by multiple impurities, as well as the Anderson localization effect, can be neglected.
Under this condition, the simple relaxation-time approximation can be employed, and
the probability-conserved Boltzmann equation reads~\cite{Blanter,Sheng}
\begin{equation}\label{Origin}
\frac{\partial f}{\partial t}+\mathbf{v}\cdot\frac{\partial f}{\partial\mathbf{r}}+\mathbf{F}\cdot
\frac{\partial f}{\partial\mathbf{k}}=-\frac{f-\langle f\rangle}{\tau_0}\ ,
\end{equation}
where $\mathbf{F}$ is the external force,
and $\mathbf{v}=\frac{\partial \varepsilon_{\mathbf{k}}}{\partial \mathbf{k}}$ is the velocity of the electron
with $\varepsilon_{\mathbf{k}}=\frac{k^2}{2m_{e}}$ as the electron energy. The relaxation time $\tau_0$
due to impurity scattering is taken to be independent of $\mathbf{k}$
and $\mathbf{r}$. $\langle f\rangle$ stands for the angular momentum average of $f(\mathbf{k},\mathbf{r})$.
For example, in three dimension,
by representing the momentum in
a polar coordinate system $\mathbf{k}=(k,\theta,\varphi)$, the angular momentum average
can be expressed as $\langle f\rangle=\frac{1}{4\pi}
\int_{0}^{\pi} d\theta \sin\theta\int_{0}^{2\pi}d\varphi f(\mathbf{k},{\bf r})$.
A unified expression of $\langle\cdots\rangle$ suitable for all $D=1,2,3$ dimensions
will be given later.
By integrating over the momentum on 
the both sides of the Boltzmann equation Eq.\ (\ref{Origin}),
one can find that the conservation law of probability is always satisfied.

\begin{figure}
\begin{center}
\includegraphics[width=7.0cm]{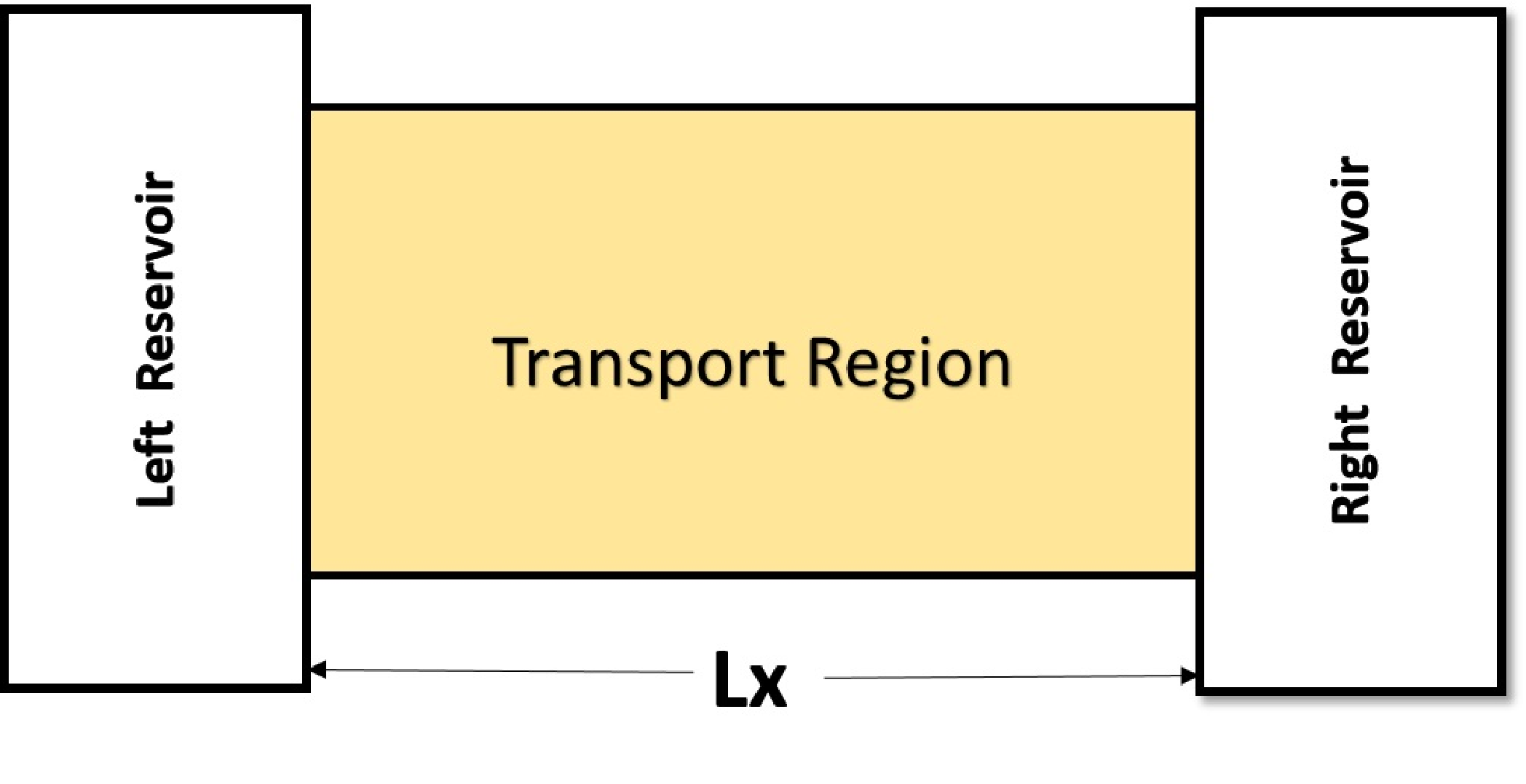}
\caption{Schematic of the setup under consideration. A finite-size conductor (transport region) is
connected to two large reservoirs, which serve as drain and source of the electrical
current. }\label{Model}
\end{center}
\end{figure}

We assume that the conductor is connected to two large reservoirs
in regions $x<0$ and $x>L_{x}$, respectively, as shown in Fig.\ \ref{Model}.
An electric field ${\bf E}(x)=E(x)\hat{\bf e}_{x}$,
with $\hat{\bf e}_{x}$ as a unit vector along the $x$
direction, is applied across the sample, and so ${\bf F}=e{\bf E}(x)$.
The electric field is confined
in the conductor, and satisfies the constraint
$V_{\mathrm{L}}-V_{\mathrm{R}}=\int_{0}^{L_{x}}E(x)dx$
with $V_{\mathrm{L}}$ and $V_{\mathrm{R}}$ being the electrical voltages
at the left and right ends of the sample.
The reservoirs remain in equilibrium, and serve as
source and drain of the electrical current, so that the electrical current can flow
through the sample continuously. The concrete position dependence of ${\bf E}(x)$
depends on the nonequilibrium charge distribution in the sample. However, we
will show that as
far as the electrical current is concerned, the result is independent of
the profile of ${\bf E}(x)$.

To proceed, it is considered that a stationary transport state has been established,
so that $\frac{\partial f}{\partial t}=0$. The electric field is taken to be small,
and we linearize the Boltzmann equation, by writing
\begin{equation}
f(\mathbf{k},x)=f_0+\left(-\frac{\partial f_0}
{\partial \varepsilon_{\mathbf{k}}}\right)w(\mathbf{k},x)\ ,\nonumber
\end{equation}
where $f_{0}=
\frac{1}{e^{(\varepsilon_{\mathbf{k}}-\varepsilon_{\mathrm{F}})/k_{\mathrm{B}}T}+1}$
is the equilibrium Fermi distribution function of the electrons.
To the linear order in $E(x)$, the Boltzmann equation reads
\begin{equation}\label{Bquation1}
v_x \frac{\partial{w}}{\partial{x}}-v_xeE(x)
=-\frac{w-\langle w\rangle}{\tau_0}\ .
\end{equation}
We notice that at low temperatures, $\left(-\frac{\partial f_0}
{\partial \varepsilon_{\mathbf{k}}}\right)$ is a delta function.
Therefore, for all $D=1$, $2$, and $3$ dimensions, the angular 
momentum average of $\langle\cdots\rangle$
can be expressed in a unified form
\begin{equation}\label{LnBoltz}
\left\langle \dots \right\rangle=\frac{\int \dots \left(  - \frac{\partial{f_0}}
{\partial{\varepsilon_{\mathbf{k}}}} \right)
d^{\mbox{\tiny{$D$}}}k}{\int\left(-\frac{\partial{f_0}}{\partial{\varepsilon_{\mathbf{k}}}} \right)d^{\mbox{\tiny{$D$}}}k}\ .
\end{equation}
For right-moving electrons $(v_{x}>0)$,
when they just move across the left interface at $x=0$ from the left reservoir into the sample,
their distribution function should still be in the equilibrium state, as they
have not been accelerated by the electric field. As a result, a boundary condition
at the left interface can be written as
\begin{equation}\label{Bcond11}
w(\mathbf{k},x=0^{+})=0\mbox{ }\hspace{1cm}(v_{x}>0)\ .\\
\end{equation}
For left-moving electrons ($v_{x}<0$), a similar boundary condition exists
at the right interface
\begin{equation}
\label{Bcond12}
w(\mathbf{k},x=L_{x}-0^{+})=0\mbox{ }\hspace{1cm}(v_{x}<0)\ .
\end{equation}

We can eliminate the electric field in the Boltzmann equation,
using the following transformation
\begin{equation}\label{transform}
w({\bf k}, x)=g({\bf k}, x)-eV_{\mathrm{L}}+e\int_{0}^{x}E(\xi)d\xi\ .
\end{equation}
By substitution of Eq.\ (\ref{transform}),
we derive Eqs.\ (\ref{Bquation1}), (\ref{Bcond11}) and  (\ref{Bcond12})
to be
\begin{equation}\label{Bquation2}
v_x \frac{\partial{g}}{\partial{x}}
=-\frac{g-\langle g\rangle}{\tau_0}\ ,
\end{equation}
and
\begin{align}\label{Bcond21}
&g(\mathbf{k},x=0^{+})=eV_{\mathrm{L}}\equiv g_{\mathrm{L}}&\mbox{ }(v_{x}>0)\ ,\\
\label{Bcond22}
&g(\mathbf{k},x=L_{x}-0^{+})=eV_{\mathrm{R}}\equiv g_{\mathrm{R}}&\mbox{ }(v_{x}<0)\ .
\end{align}
The electric field
$E(x)$ with its profile unknown no longer appears in the Boltzmann equation
Eq.\ (\ref{Bquation2}), and instead, electrical voltages
$V_{\mathrm{L}}$ and $V_{\mathrm{R}}$
appear in the boundary conditions Eqs.\
(\ref{Bcond21}) and  (\ref{Bcond22}).
It is easy to show that the
expression for the electrical current density ${\bf j}$ is invariant
under the above transformation, i.e.,
\begin{eqnarray}
{\bf j}&=&\frac{2e}{h^{\mathrm{D}}}\int{{\bf v}w\left(\mathbf{k},x\right)
\left(-\frac{\partial f_0}{\partial \varepsilon_{\mathbf{k}}}\right)d^{\mbox{\tiny{$D$}}}k}
\nonumber\\
&\equiv& \frac{2e}{h^{\mathrm{D}}}\int{{\bf v}g\left(\mathbf{k},x\right)
\left(-\frac{\partial f_0}{\partial \varepsilon_{\mathbf{k}}}\right)d^{\mbox{\tiny{$D$}}}k}\ .
\nonumber
\end{eqnarray}
Therefore, $g({\bf k},x)$ plays the same role as $w({\bf k},x)$ does.
We point out that the transformation Eq.\ (\ref{transform}) is 
valid in the linear regime, and only in this regime, 
gauge invariance of the physical quantities 
calculated from Eqs.\ (\ref{Bquation2})-(\ref{Bcond22}),
with respect to different choices of the electrical voltages $V_{\mathrm{L}}$
and $V_{\mathrm{R}}$, is guaranteed.
The above derivation proves that the electrical current depends only on the electrical voltage
difference across the sample, independent of the profile of the electric field
$E(x)$. Owing to this finding, laborious calculations of the electric field
from the Maxwell's equations are avoided.

We note that $\langle g\rangle$ is a function of coordinate $x$
only, which will be denoted as
$\left\langle g\right\rangle\equiv\bar{g}(x)$ for clarity in the following formulation.
According to the definition Eq.\ (\ref{transform}),
$\bar{g}(x)$ describes both the effects of the applied electric field
and accumulation of carriers $\langle f\rangle-f_{0}$. From the boundary conditions
Eqs.\ (\ref{Bcond21}) and  (\ref{Bcond22}), $\bar{g}(x)$ has the same unit as the
chemical potential. Besides, since electric field no longer appears in the Boltzmann equation,
and it is the gradient of $\bar{g}(x)$, which drives the
electrical current. Therefore, we may call $\bar{g}(x)$ the effective
chemical potential (more strictly, change in chemical potential induced
by the applied electric field). 
Eq.\ (\ref{transform}) may be regarded as a transformation
from the representation of charged particles, where the driving force for transport
is the electric field, to a representation of neutral particles, where 
the gradient of the chemical potential causes flow of the particles.

\section{The exact solution}\label{s21}

\begin{figure}
\begin{center}
\includegraphics[width=7.0cm]{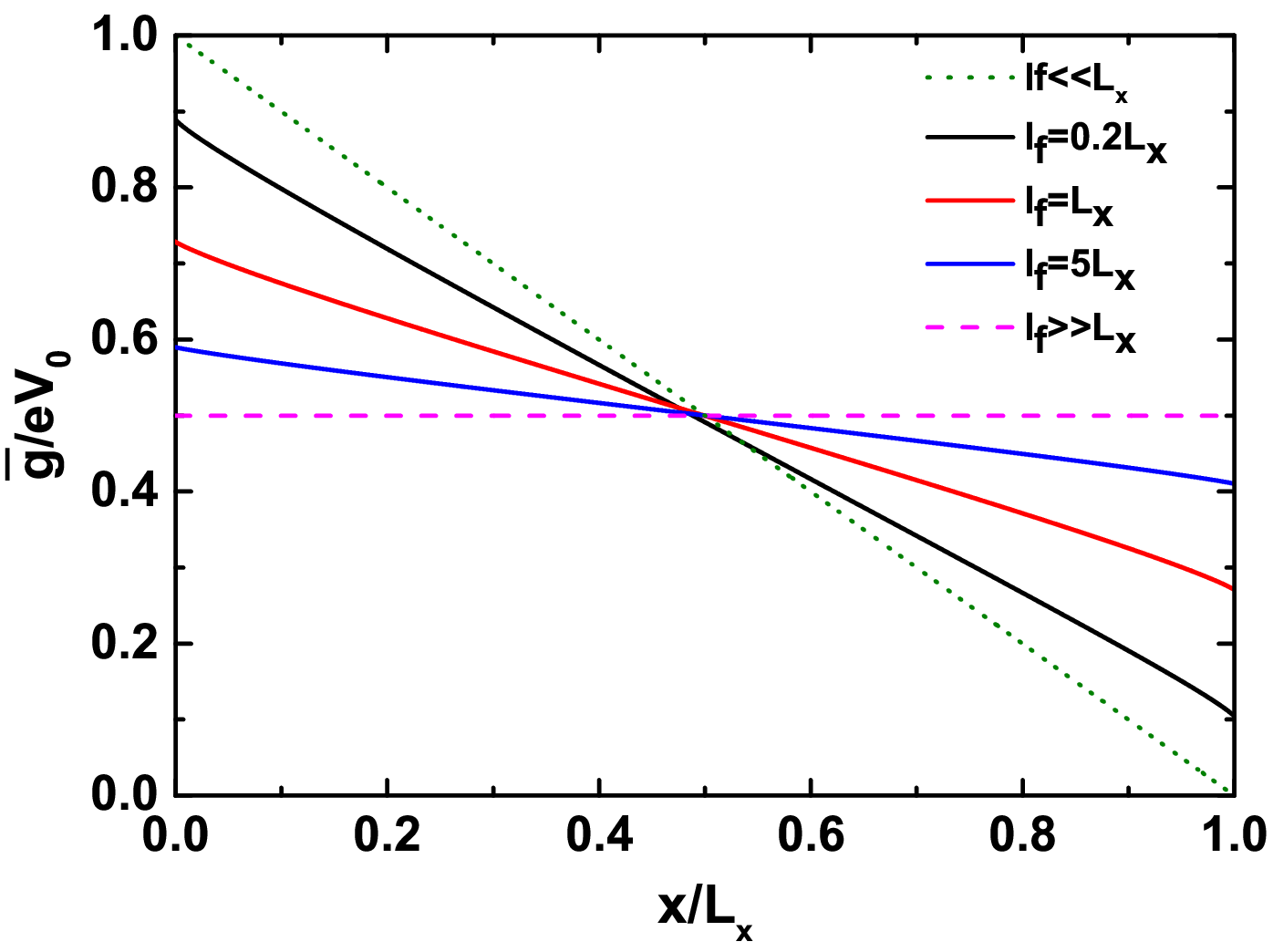}
\caption{Exact numerical solution of the effective chemical pontential
$\bar{g}(x)$ in two dimension as a function of $x/L_{x}$ for some different
mean free path to sample length ratios. Here, we choose $V_{\mathrm{L}}=V_0$,
and $V_{\mathrm{R}}=0$.}\label{gbarfig}
\end{center}
\end{figure}
A formal solution of $g({\bf k},x)$ can be obtained from the Boltzmann equation Eq.\ (\ref{Bquation2}) and  boundary conditions Eqs.\ (\ref{Bcond21}) and  (\ref{Bcond22}), as a linear functional
of $\bar{g}(x)$
\begin{widetext}
\begin{eqnarray}\label{g}
g({\bf k},x)=\theta(v_x)\left(g_\mathrm{L} e^{-\frac{x}{v_x \tau_0}}
+\int_{0}^{x}e^{-\frac{x-\xi}{v_x\tau_0} }\bar{g}(\xi)\frac{d\xi}{{v_x\tau_0}}\right)
+\theta(-v_x)\left(g_\mathrm{R} e^{-\frac{x-L_x}{v_x \tau_0}}+ \int_{L_x}^{x}e^{-\frac{x-\xi}{v_x\tau_0} }\bar{g}(\xi) \frac{d\xi}{v_x\tau_0} \right)\ ,
\end{eqnarray}
where $\theta(v_{x})$ is the unit step function.
By taking the angular momentum average on the both sides 
of Eq.\ (\ref{g}), we obtain a self-consistent integral equation
for $\bar{g}(x)$
\begin{eqnarray}\label{barg}
h(x)=\bar{g}(x)-\int_{0}^{x} {K_1(x,\xi)\bar{g}(\xi)d\xi}
-\int_{L_x}^{x} {K_2(x,\xi)\bar{g}(\xi)d\xi}\ ,
\end{eqnarray}
\end{widetext}
where
\begin{align*}
&h(x)= \left\langle\theta\left(v_x\right) g_\mathrm{L} e^{-\frac{x}{v_x \tau_0}} \right\rangle+\left\langle\theta\left(-v_x\right) g_\mathrm{R} e^{-\frac{x-L_x}{v_x \tau_0}} \right\rangle\ ,\\
&K_1(x,\xi)=\left \langle \theta\left(v_x\right)\frac {1}{v_x\tau_0} e^{-\frac{x-\xi}{v_x\tau_0} }\right \rangle\ ,\\
&K_2(x,\xi)=\left \langle \theta\left(-v_x\right) \frac {1}{v_x\tau_0}e^{-\frac{x-\xi}{v_x\tau_0} }
\right\rangle\ .
\end{align*}

Equations (\ref{g}) and (\ref{barg}) constitute an exact solution to the Boltzmann equation
Eq.\ (\ref{Bquation2}).
One can solve the effective chemical potential field $\bar{g}(x)$  from Eq.\ (\ref{barg}), and then
substitute it into Eq.\ (\ref{g}) to obtain the nonequilibrium distribution function $g({\bf k},x)$.
Once $g({\bf k},x)$ is obtained, the nonequilibrium transport properties of the
system can be determined. In general, it is not easy to obtain the exact analytical
expression for $\bar{g}(x)$. Before we work out an approximate analytical
solution in the next section, we now carry out numerical calculation.
Through discretization of the coordinate $x\in [0,L_{x}]$,
Eq.\ (\ref{barg}) is reduced to a set of linear equations, which can be solved
numerically.
The calculated effective chemical potential field $\bar{g}(x)$ for a two-dimensional sample
is plotted in Fig.\ \ref{gbarfig} as a function of $x/L_{x}$
for several different mean free path to sample length ratios. The mean free path
is defined as $l_{f}=v_{\mathrm{F}}\tau_{0}$ with $v_{\mathrm{F}}$ being
the Fermi velocity.
From Fig.\ \ref{gbarfig}, we see that $\bar{g}(x)$ is exactly a linear function of $x$
in the two limits $L_{x}\ll l_f$ and $L_{x}\gg l_f$. In fact, it is easy to
obtain from Eq.\ (\ref{barg})
\begin{equation}\label{g_lim}
\bar{g}(x)=
\begin{cases}
g_\mathrm{L}-(g_{\mathrm{L}}-g_{\mathrm{R}})x/L_{x}&\mbox{ }L_{x}\gg l_f\\
(g_{\mathrm{L}}+g_{\mathrm{R}})/2&\mbox{ }L_{x}\ll l_f
\end{cases}
\ .
\end{equation}
When $L_{x}$ is comparable to $l_f$, a linear dependence
is still valid in the middle region of the sample, 
but tiny deviations from the linear dependence
occur near the sample boundaries $x=0$ and $x=L_{x}$.

\section{An Analytical Approximation}\label{s3}
As has been observed in Sec.\ \ref{s21}, the exact solution of
the effective chemical potential $\bar{g}(x)$ is nearly a linear function
of coordinate $x$ with negligible deviations occurring
near the sample boundaries in the region $L_{x}\sim l_{f}$. Therefore,
it is reasonable to make a linear approximation to $\bar{g}(x)$, assuming
\begin{equation}\label{gbarl}
\bar{g}(x)=a+bx\ ,
\end{equation}
with $a$ and $b$
as two constant coefficients to be determined. Substituting this
trial solution into Eq. (\ref{barg}), we get an equation for $a$ and $b$
\begin{equation}\label{par}
\begin{split}
0=&\left\langle\theta\left(v_x\right)\left(g_\mathrm{L}-a+bv_x\tau_0\right)e^{-\frac{x}{v_x\tau_0}} \right\rangle+\\
&\left\langle\theta\left(-v_x\right)\left(g_\mathrm{R}-a-bL_x+bv_x\tau_0\right)e^{-\frac{x-L_x}{v_x\tau_0}} \right\rangle\ .
\end{split}
\end{equation}
To determine the coefficients $a$ and $b$, one can choose two different values of coordinate $x$
in the above equation to obtain
a couple of equations of $a$ and $b$. Noticing that the linear dependence of $\bar{g}(x)$ on
$x$ is very well satisfied in the middle region of the sample, we choose
$x=\frac{L_x}{2}$ and $x=\frac{L_x}{2}+\Delta x$, and take the limit $\Delta x\rightarrow 0$
in the final solution. We obtain
\begin{align}\label{a_expr}
a&=\frac{L_{x}g_\mathrm{L}+\kappa l_{f}(g_\mathrm{{L}}+g_\mathrm{R})}{L_{x}+2\kappa l_{f}}\ ,\\
\label{b_expr}
b&=-\frac{g_\mathrm{L}-g_\mathrm{R}}{L_x+2\kappa l_f}\ ,
\end{align}
where
\begin{equation}\label{kappa}
\kappa=\frac{\tau_{0}\left\langle \theta(v_x)e^{-\frac{L_{x}}{2v_{x}\tau_{0}}}\right\rangle}
{\l_{f}\left\langle \theta(v_x)\frac{1}{v_{x}}e^{-\frac{L_{x}}{2v_{x}\tau_{0}}}\right\rangle}\ .
\end{equation}
Notebaly, Eq.\ (\ref{gbarl}) together with Eqs.\ (\ref{a_expr}) and (\ref{b_expr}) 
recover Eq.\ (\ref{g_lim}) in the two limits $L_{x}\gg l_{f}$ and $L_{x}\ll l_{f}$. 

%\begin{equation}\label{Part}
%2a+bL_x=g_{\mathrm{L}}+g_{\mathrm{R}}~\ .
%\end{equation}

\subsection{One Dimension}
For $D=1$ dimension, in Eq.\ (\ref{kappa}) $v_{x}\equiv v_{\mathrm{F}}$, and
so 
\begin{equation}
\kappa\equiv 1\ .
\end{equation}
Interestingly, we notice that if we substitute Eqs.\ (\ref{a_expr}) and (\ref{b_expr})
with $\kappa=1$ into
Eq.\ (\ref{par}), both terms on the right-hand side of Eq.\ (\ref{par}) vanish
identically for any $x$.
This means that for $D=1$, Eq.\ (\ref{gbarl}) is actually an exact solution to Eq.\ (\ref{barg}). 
Therefore, the conductance formula obtained below for one-dimensional systems is an
exact result of the Boltzmann equation.
Since the electrical current $I$ is constant along the $x$ direction, we
calculate $I$ setting $x=L_{x}/2$, yielding
\begin{equation}
\begin{split}\label{Curr}
I&=\frac{2e}{h}\int{v_xg\left(k_x,x=\frac{L_{x}}{2}\right)\left(-\frac{\partial f_0}{\partial \varepsilon_{k}}\right)dk_x}\ \\
&=G^{1\mathrm{D}}(V_{\mathrm{L}}-V_{\mathrm{R}})\ ,\nonumber
\end{split}
\end{equation}
where
\begin{equation}\label{G_D1}
G^{1\mathrm{D}}=G_0^{1\mathrm{D}}\frac{2l_f}{L_x+2l_f}\ ,
\end{equation}
is the conductance of the system. Here, $G_0^{1\mathrm{D}}=N_{\mathrm{ch}}\frac{e^2}{h}$,
with $N_{\mathrm{ch}}=2$
taking into account the spin degeneracy. We note that the electrical current $I$ depends only on 
the voltage difference $(V_{\mathrm{L}}-V_{\mathrm{R}})$ between the two ends of the 
sample, which is a manifestation of the gauge invariance. 
In the ballistic limit $L_{x}\ll l_{f}$, $G^{1\mathrm{D}}= G_0^{1\mathrm{D}}$, being consistent
with the Landauer-B\"{u}ttiker formula. In the diffusive limit $L_{x}\gg l_{f}$,
$G^{1\mathrm{D}}=\frac{1}{L_{x}}\frac{n_{e}e^2\tau_{0}}{m_{e}}$
with $n_{e}=\frac{4k_{\mathrm{F}}}{h}$ as the electron density, which recovers the
well-known Boltzmann-Drude formula.  

\begin{figure}
\begin{center}
\includegraphics[width=8.0cm]{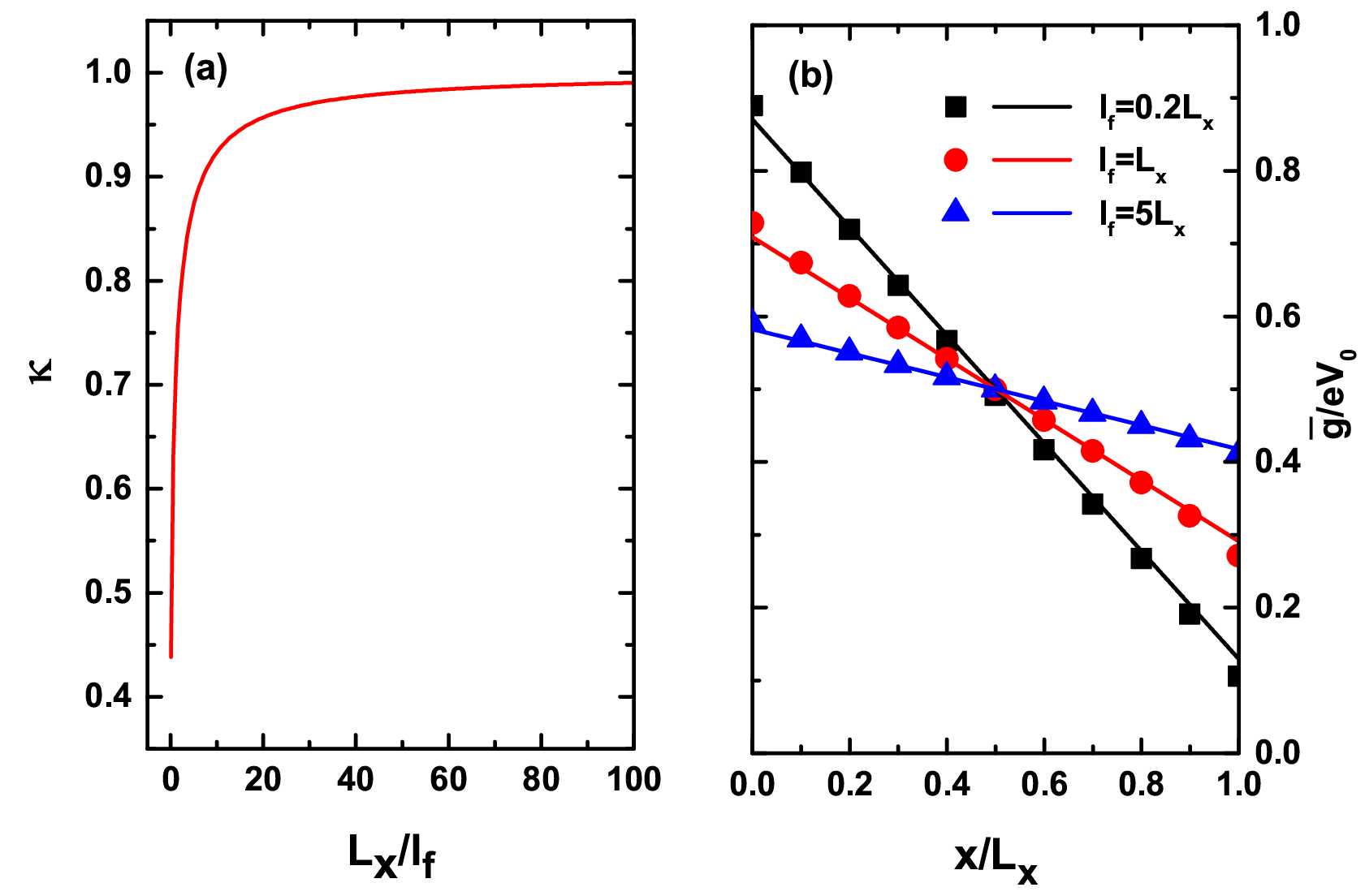}
\caption{(a) The parameter $\kappa$ as a function of $L_x/l_f$ in two dimension.
(b) The linear approximation to $\bar{g}(x)$ (solid lines), compared with the exact
numerical solution reproduced from Fig.\ \ref{gbarfig} (symbols).
}\label{gbar2fig}
\end{center}
\end{figure}

\subsection{Two Dimension}
In $D=2$ dimension, by using a polar coordinate system, we obtain
\begin{equation}\label{kappa2}
\kappa=\frac{\int_{-\frac{\pi}{2}}^{\frac{\pi}{2}} e^{-\frac{L_x}{2l_f \cos\phi}}d\phi}{\int_{-\frac{\pi}{2}}^{\frac{\pi}{2}} \frac{1}{\cos\phi}e^{-\frac{L_x}{2l_f \cos\phi}}d\phi}\ .
\end{equation}
The calculated $\kappa$ is plotted in Fig.\ \ref{gbar2fig}(a) as a function of $L_{x}/l_{f}$.
With increasing $L_{x}/l_{f}$ from $0$, $\kappa$ increases from $0$ and then
approaches $1$ rapidly. In Fig.\ \ref{gbar2fig}(b), the approximate solution
$\bar{g}(x)=a+bx$, with $a$ and $b$ given by Eqs.\ (\ref{a_expr}) and (\ref{b_expr}),
is plotted  as solid lines for some different values of $L_{x}/l_{f}$.
The exact solution of $\bar{g}(x)$ is also shown as symbols in Fig.\ \ref{gbar2fig}(b).
The approximate solution fits very well with the exact solution.
The electrical current at $x=\frac{L_x}{2}$ is calculated, yielding
\begin{equation}
\begin{split}
I&=\frac{2eL_{y}}{h^2}\int v_xg\left({\bf k},x=\frac{L_x}{2}\right)
\left(-\frac{\partial f_0}{\partial\varepsilon_{\mathbf{k}}}\right)dk_xdk_y\\
&=G^{2\mathrm{D}}
(V_{\mathrm{L}}-V_{\mathrm{R}})\ ,\nonumber
 %\int_{-\frac{\pi}{2}}^{\frac{\pi}{2}}  d\phi
%-bv_x\tau_0+\theta(v_x)\left(g_L-a+bv_x\tau_0\right)e^{-\frac{x}{v_x\tau_0}}\\
%&+\theta(-v_x)\left(g_R-a-bL_x+bv_x\tau_0\right)e^{-\frac{x-L_x}{v_x\tau_0}}
\end{split}
\end{equation}
where
\begin{equation}\label{G_2D}
G^{2\mathrm{D}}={G}_0^{2\mathrm{D}}(\chi^{2\mathrm{D}}_{\mathrm{bal}}+\chi^{2\mathrm{D}}_{\mathrm{dif}})
\end{equation}
is the conductance of the two-dimensional sample. Here,
$G_0^{2\mathrm{D}}=N_{\mathrm{ch}}\frac{e^2}{h}$, where $N_{\mathrm{ch}}=\frac{4k_{\mathrm{F}}L_{y}}{h}$
is the channel number with $L_{y}$ as the cross-section length
of the sample, and
\begin{align}
\chi^{\mathrm{2D}}_{\mathrm{bal}}=&\frac{\kappa l_f}{L_x+2\kappa l_f}\int_{-\frac{\pi}{2}}^{\frac{\pi}{2}} e^{-\frac{L_x}{2l_f\cos\phi}}\cos\phi d\phi\ ,\nonumber \\
\chi^{\mathrm{2D}}_{\mathrm{dif}}=&\frac{ l_f}{L_x+2\kappa l_f}\int_{-\frac{\pi}{2}}^{\frac{\pi}{2}} \left(1-e^{-\frac{L_x}{2l_f\cos\phi}}\right)\cos^2\phi d\phi\ .\nonumber
\end{align}
The total conductance is divided into two parts:
${G}_0^{2\mathrm{D}}\chi^{2\mathrm{D}}_{\mathrm{bal}}$
and ${G}_0^{2\mathrm{D}}\chi^{2\mathrm{D}}_{\mathrm{dif}}$,
standing for
contributions from electron ballistic and diffusive transport
processes, respectively.
In the ballistic limit $L_{x}\ll l_{f}$, $\chi_{\mathrm{bal}}
\rightarrow 1$ and $\chi_{\mathrm{dif}}\rightarrow 0$,
such that $G^{2\mathrm{D}}=G_0^{2\mathrm{D}}$, being consistent
with the Landauer-B\"{u}ttiker formula. In the diffusive limit $L_{x}\gg l_{f}$,
$\chi_{\mathrm{bal}}\rightarrow 0$ and $\chi_{\mathrm{dif}}\rightarrow \frac{\pi l_{f}}{2L_{x}}$,
and so $G^{2\mathrm{D}}=\frac{L_{y}}{L_{x}}\frac{n_{e}e^2\tau_{0}}{m_{e}}$,
with $n_{e}=\frac{2\pi k^2_{\mathrm{F}}}{h^2}$ as the electron density, recovers the
Boltzmann-Drude formula.

%What's more we can use some approximation in the integrations in $\chi_{\dots}$ which is just take $\cos\phi$ in %$e^{-L_x/(2l_f\cos\phi)}$ as constant $1$ .
%Finally we can get result of $\chi_{\dots} $ as
%\begin{align}
%\chi_{\mathrm{bal}}=&\frac{2 l_f}{L_x+2 l_f}e^{-\frac{L_x}{2l_f}} \\
%\chi_{\mathrm{cla}}=&\frac{ \pi l_f}{2(L_x+2\kappa l_f)}\left(1-e^{-\frac{L_x}{2l_f}}\right)~.
%\end{align}
%We can do this because for both $\chi_{bal}$ and $\chi_{cla}$,when $\phi$ satisfies $\cos\phi\ll 1$ the integration %is small compared with $\cos\phi=1$.

\subsection{Three Dimension}
In $D=3$ dimension, we obtain for $\kappa$
\begin{equation}\label{kappa3D}
\kappa=\frac{\int_{0}^{1} e^{-\frac{L_x}{2l_f u}}
du}{\int_{0}^{1}\frac{1}{u} e^{-\frac{L_x}{2l_f u}}
du}\ .
\end{equation}
The conductance is derived to be
\begin{equation}\label{G_3D}
{G}^{3\mathrm{D}}={G}_0^{3\mathrm{D}}(\chi_{\mathrm{bal}}^{3\mathrm{D}}+\chi_{\mathrm{dif}}^{3\mathrm{D}})
\end{equation}
where ${G}_0^{3\mathrm{D}}=N_{\mathrm{ch}}\frac{e^2}{h}$ with $N_{\mathrm{ch}}=\frac{2\pi k_{\mathrm{F}}^2}{h^2}A$
with $A$ as the cross section,
\begin{align}
\chi^{\mathrm{3D}}_{\mathrm{bal}}=&\frac{4\kappa l_f}{L_x+2\kappa l_f}\int_{0}^{1} e^{-\frac{L_x}{2l_fu}}u du\ ,
\nonumber \\
\chi^{\mathrm{3D}}_{\mathrm{dif}}=&\frac{4l_f}{L_x+2\kappa l_f}\int_{0}^{1}
\left(1-e^{-\frac{L_x}{2l_fu}}\right)u^2 du\ .\nonumber
\end{align}
In the ballistic limit $L_{x}\ll l_{f}$, $\chi_{\mathrm{bal}}\rightarrow 1$ and $\chi_{\mathrm{dif}}\rightarrow 0$,
and hence $G^{3\mathrm{D}}=G_0^{3\mathrm{D}}$, in agreement with
the Landauer-B\"{u}ttiker formula. In the diffusive limit $L_{x}\gg l_{f}$,
$\chi_{\mathrm{bal}}\rightarrow 0$ and $\chi_{\mathrm{dif}}\rightarrow \frac{4 l_{f}}{3L_{x}}$.
As a result, $G^{3\mathrm{D}}=\frac{A}{L_{x}}\frac{n_{e}e^2\tau_{0}}{m_{e}}$,
with $n_{e}=\frac{8\pi k^3_{\mathrm{F}}}{3h^3}$ as the electron density, reproduces the
Boltzmann-Drude formula.\\

\begin{figure}
\begin{center}
\includegraphics[width=6.7cm]{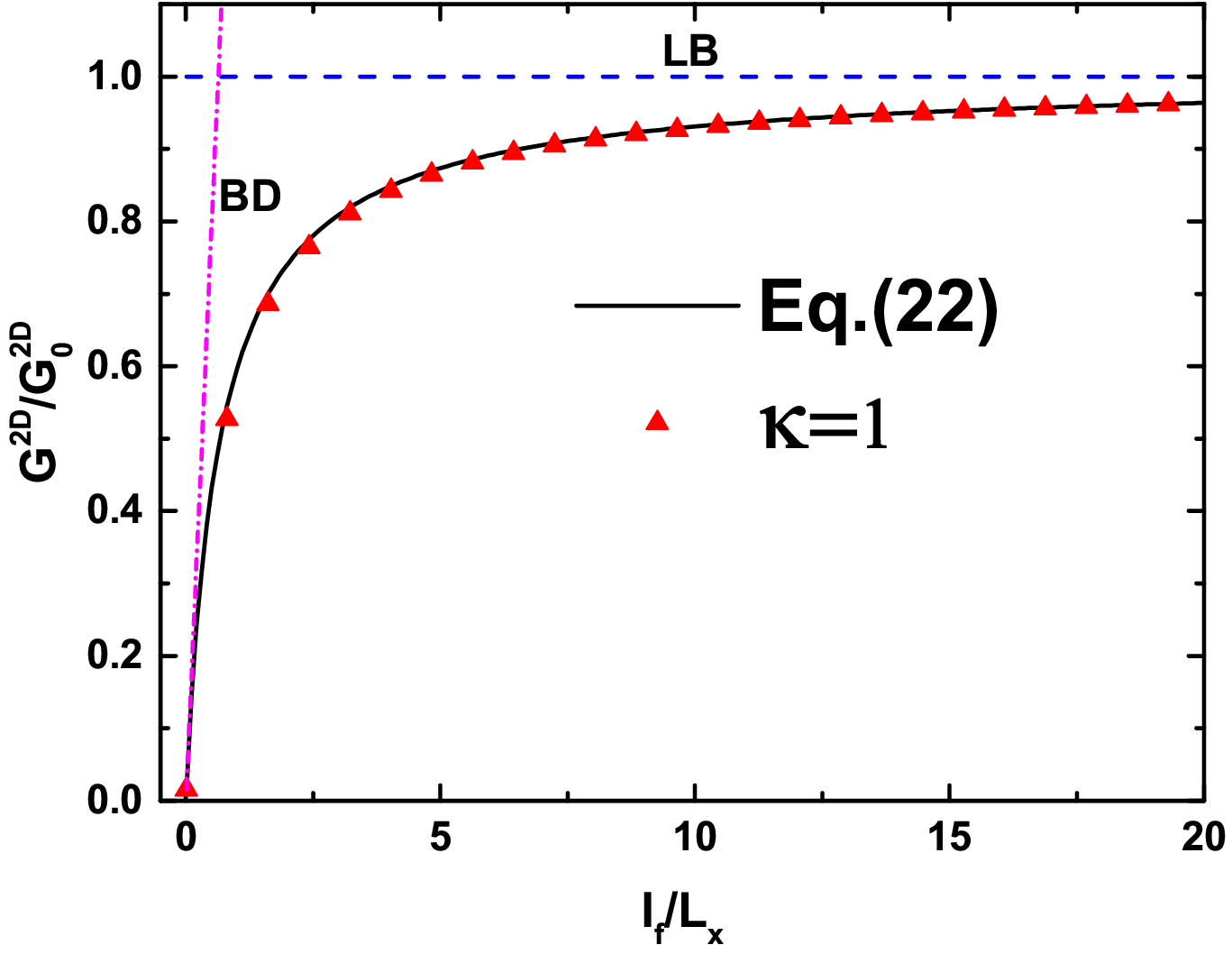}
\caption{Conductance of a two-dimensional conductor as a function of
$l_{f}/L_{x}$ calculated by using the expression Eq.\ (\ref{kappa2}) for $\kappa$ (solid line),
compared with that calculated by setting $\kappa=1$ (triangles).
Dot-dashed line represents the Boltzmann-Drude (BD) formula $G=\frac{n_{e}e^2\tau_{0}}{m_{e}}
\frac{L_{y}}{L_{x}}$, and the dashed line stands for
the Landauer-B\"{u}ttiker (LB) formula $G=N_{\mathrm{ch}}\frac{e^2}{h}$.}\label{conductance}\label{CondFig}
\end{center}
\end{figure}

We wish to point out that for both $D=2$ and $3$, a further simplification of
the theory can be done
by setting $\kappa=1$. From Fig.\ \ref{gbar2fig}(b), we
see that $\kappa$ has large deviations from $1$ for $L_{x}\lesssim l_{f}$.
Fortunately, as can be seen from Eqs.\ (\ref{G_2D}) and (\ref{G_3D}),
the conductance becomes insensitive to the value
of $\kappa$ in the region $L_{x}\lesssim l_{f}$. In Fig.\ \ref{CondFig},
the conductance for $D=2$ calculated by using $\kappa=1$ is compared with
that obtained by using the expression Eq.\ (\ref{kappa2}) for $\kappa$.
The difference between them is nearly invisible, an indication that
setting $\kappa=1$
is a very good approximation for most purposes.

\section{A Summary and Discussions}\label{s4}

In summary, we have demonstrated that the Boltzmann equation together with proper boundary conditions
can describe electron transport properties from diffusive to ballistic regimes.
We have worked out a sufficiently accurate analytical solution to the Boltzmann
equation. Analytical formulas of the electrical conductance for $D=1$, $2$, and $3$ dimensions are
obtained, which smoothly bridge the Boltzmann-Drude formula in diffusive regime and
Landauer B\"{u}ttiker formula in ballistic regime.
This simple and intuitive approach can be applied to investigate the effects of
sample size and impurity scattering in various charge transport phenomena.

Spin-dependent electronic transport has attracted a great deal of interest in recent years.
To describe spin-dependent transport, some essential generalizations of
the present theory need to be done. First, a spin relaxation mechanism needs to 
be included. Second,
the effective chemical potential field $\bar{g}({\bf r})$ defined 
in the present theory includes the effects of
the applied electric field and accumulation of particles. In a spin-dependent
transport process, while the physical electric field is always spin-independent,
the accumulation of particles can become spin-dependent, resulting
in the so-called spin accumulation. Therefore,
a spin-dependent effective chemical potential field $\bar{g}_{s}({\bf r})$ can be
assumed to account for the effect of spin accumulation.

\begin{acknowledgments}
This work was supported by the State Key Program for Basic Researches of China under
grants numbers 2015CB921202 and 2014CB921103, the National Natural Science Foundation of China under grant numbers 11225420,
 and a project funded by the PAPD of Jiangsu Higher Education Institutions.
\end{acknowledgments}

\end{document}